\documentclass[twocolumn,showpacs,preprintnumbers,amsmath,amssymb,superscriptaddress]{revtex4}

\usepackage{graphicx}% Include figure files
\usepackage{dcolumn}% Align table columns on decimal point
\usepackage{bm}% bold math

\begin{document}

\preprint{JETP Letters 81, 10 (2005)}

\title{Anomalous Behavior near $T_c$ and Synchronization
of Andreev Reflection in Two-Dimensional Arrays of SNS Junctions}

\author{T.I. Baturina} \email{tatbat@isp.nsc.ru}
\affiliation{Institute of Semiconductor Physics, Siberian Division, 
Russian Academy of Sciences, Novosibirsk, 630090 Russia}
\author{Yu.A. Tsaplin}
\affiliation{Institute of Semiconductor Physics, Siberian Division, 
Russian Academy of Sciences, Novosibirsk, 630090 Russia}
\affiliation{Novosibirsk State University, Novosibirsk, 630090 Russia}
\author{A.E. Plotnikov}
\affiliation{Institute of Semiconductor Physics, Siberian Division, 
Russian Academy of Sciences, Novosibirsk, 630090 Russia}
\author{M.R. Baklanov}
\affiliation{Interuniversity Microelectronics Center, B-3001 Leuven, Belgium}

\date{\today}

\begin{abstract}
We have investigated low-temperature transport properties 
of two-dimensional arrays of superconductor--normal-metal--superconductor (SNS) 
junctions. It has been found that
in two-dimensional arrays of SNS junctions
(i) a change in the energy spectrum within an
interval of the order of the Thouless energy is observed even when the 
thermal broadening far exceeds the Thouless energy for a single SNS junction; 
(ii) the manifestation of the subharmonic energy gap structure (SGS) with high 
harmonic numbers is possible even if the energy relaxation length 
is smaller than that required for the realization of a multiple Andreev 
reflection in a single SNS junction. These results point to the 
synchronization of a great number of SNS junctions. 
A mechanism of the SGS origin in two-dimensional arrays of SNS junctions, 
involving the processes of conventional and crossed Andreev reflection, 
is proposed. 
\end{abstract}
\pacs{73.23.-b; 74.45.+c; 74.81.Fa}

\maketitle

Andreev reflection is a microscopic mechanism
responsible for the charge transport through a 
normal-metal--superconductor interface~\cite{Andreev}.
An electron-like quasiparticle in the normal metal with an energy lower
than the superconducting gap ($\Delta$) is reflected from the
boundary as a hole-like quasiparticle, while a Cooper
pair is transferred to the superconductor. 
If the normal
metal is sandwiched between two superconductors, 
there is an additional charge transfer mechanism, namely, multiple
Andreev reflection (MAR). 
This mechanism was
proposed in~\cite{KBT} to explain the subharmonic energy gap structure
(SGS) observed as dips in the differential resistance
curves at voltages $e V_n = 2\Delta /n$ ($n$ is an integer).
The MAR concept is as follows: owing to sequential
Andreev reflections from normal-metal--superconductor
(NS) interfaces, a quasiparticle passing through the
normal region can accumulate an energy of $2 \Delta$,
which is sufficient for the transition to single-particle states of
the superconductor. Although, by now, ample experimental
data on the transport properties of NS and SNS junctions are 
available~\cite{Pannetier} and the existing theoretical models
fairly well describe the phenomena observed in the
experiments~\cite{BTK,OTBK,Bez}, the properties of multiply connected
SNS systems are poorly investigated. It should
be noted that systems consisting of superconducting
islands incorporated into a normal metal are spontaneously
formed in disordered superconducting films~\cite{Q_SMT}.
In view of these circumstances, it is of interest to study the
properties of model multiply connected SNS systems with superconducting 
and normal regions formed in a controlled way.

\begin{figure}
\includegraphics[width=1. \columnwidth]{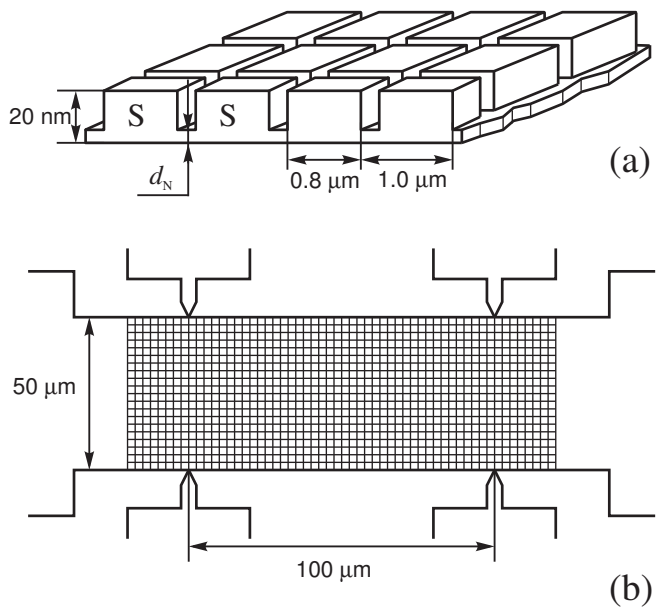}
\caption{Schematic representation of the structure: (a) profile
of the film modulated in thickness (the geometric dimensions
are indicated in the figure); (b) position of the structured
film on the sample used for the measurements.}
\label{fig1}
\end{figure}

This paper is devoted to the properties of two-dimensional
arrays of SNS junctions fabricated on the
basis of a 20-nm-thick PtSi superconducting film (with
a critical temperature of $T_c = 0.64$~K)~\cite{PtSitech}.
The transport parameters of the initial PtSi film were as follows: the
resistance per square at $T = 4.2$~K was $R_{sq} = 22.8$~$\Omega$,
the mean free path was $l= 1.35$~nm, and the diffusion coefficient was
$D = 7.2$ cm$^2$/s. The initial samples were fabricated
by photolithography in the form of Hall bars
50 $\mu$m wide and 100 $\mu$m long.
Then, by electron beam lithography with subsequent plasma
chemical etching, the initial film was thinned in preset regions.
Figure~\ref{fig1}a schematically represents the structure
under study. It consists of periodically arranged
film areas with a thickness of 20 nm (islands), between
which the film is thinned by plasma chemical etching.
The period of the structure is 1~$\mu$m, and the island
dimensions are $0.8 \times 0.8$ $\mu$m. This structure completely
covers the whole area of the Hall bar (Fig.~\ref{fig1}b).
Thus, the number of islands between the potential terminals is
$50 \times 100$.
To control the depth of etching, a reference
film was etched uniformly over its surface simultaneously
with the structure etching. The low-temperature
transport measurements were performed by a
standard four-probe low-frequency
($\sim$10~Hz) technique. An ac current was within the range $1-10$~nA.

As the experiment shows, the decrease of film
thickness by plasma chemical etching
leads to the suppression of the critical temperature.
At the same time, in the islands,
whose thickness remains equal to that of the initial
PtSi film, the suppression of $T_c$
is minor. Hence, there is a certain
temperature region within which the thinned film is in
the normal (N) state and the islands are in the superconducting
(S) state. In this temperature region, the structures
under study represent a two-dimensional array of
SNS junctions. Moreover, since the superconducting
and normal regions of the SNS junctions, fabricated in
the aforementioned way, consist of the same material,
the formation of tunnel barriers at the NS interfaces is
excluded and a high transparency of the NS interfaces
can be {\it a priori} expected.

\begin{figure}
\includegraphics[width=\linewidth]{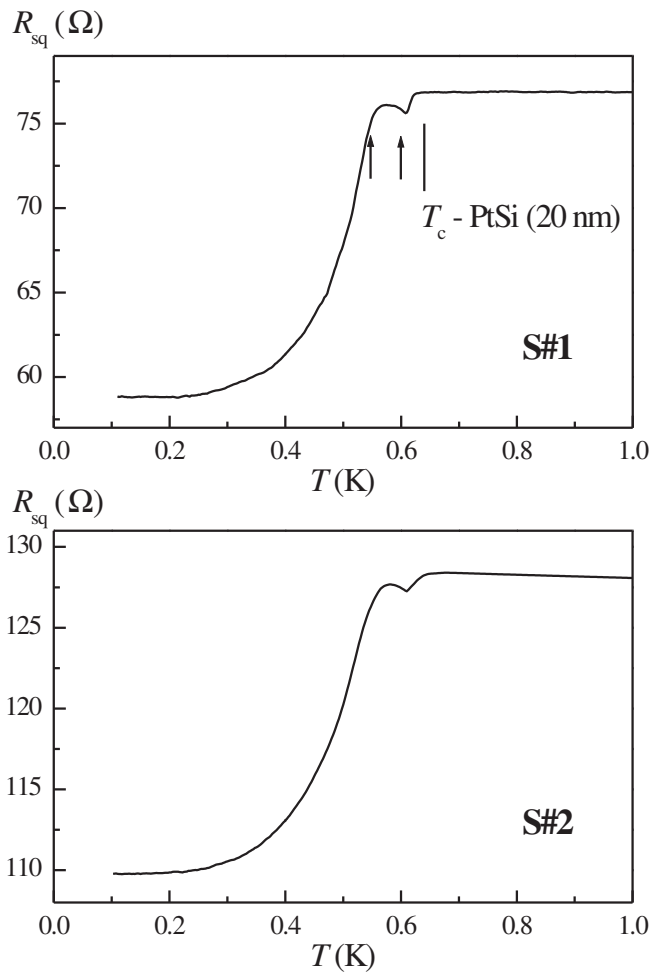}
\caption{Temperature dependence of the resistance per square
for samples {\bf S\#1} and {\bf S\#2}. The dash in the plot indicates the
superconducting transition temperature for the initial 20-nm-thick 
PtSi film. The arrows indicate the temperature
range from 0.545 to 0.599 K.}
\label{fig2}
\end{figure}

Figure~\ref{fig2} shows the temperature dependence of resistance
for two samples that only differ in the film thickness
between the islands. The resistance per square of
the reference film at $T = 4.2$~K is 995~$\Omega$
for sample {\bf S\#1} and 1483~$\Omega$ for sample {\bf S\#2}.
For all structures studied, a noticeable decrease in 
resistance with decreasing temperature is observed 
slightly below the temperature $T_c$
of the initial 20-nm-thick PtSi film. As the temperature
decreases (Fig.~\ref{fig2}), a small decrease in the resistance
is first observed at $T \sim 0.61$~K; then, the resistance
increases reaching a maximum and then decreases
again. Such an anomalous behavior of the temperature
dependence of resistance in the narrow temperature
range near $T_c$ was never observed before, and the origin
of this behavior is unknown.

\begin{figure}
\includegraphics[width=1. \columnwidth]{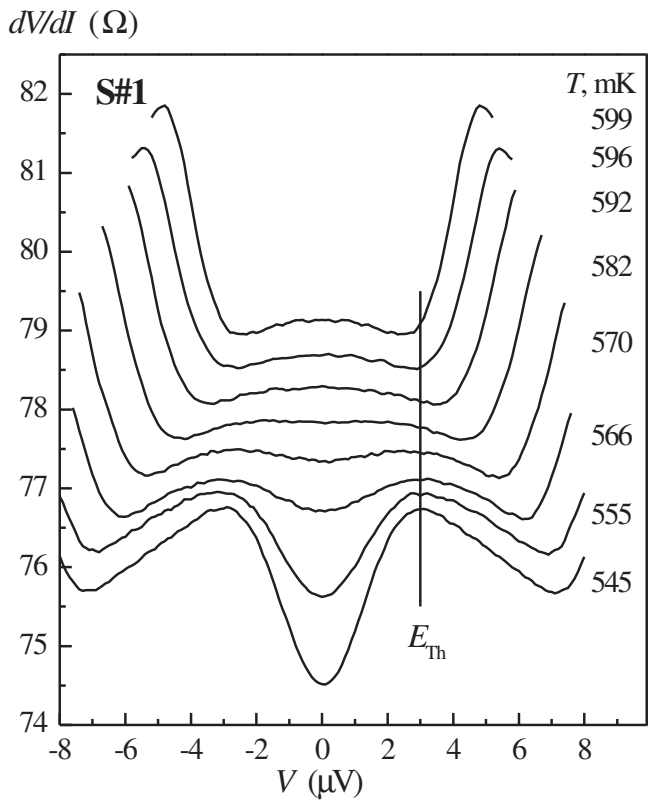}
\caption{Differential resistance per square versus the bias
voltage per SNS junction for sample {\bf S\#1}. All dependences,
except for the lower one, are sequentially shifted upwards
by $0.5 \Omega$.}
\label{fig3}
\end{figure}

To explore the nature of this anomaly, the
current--voltage characteristics of the samples have been studied. 
Figure~\ref{fig3} shows the dependences of the differential
resistance ($dV/dI$) on the bias voltage for the sample {\bf S\#1}.
The measured values of $dV/dI$ are given per square. The total
voltage is divided by the number of junctions (100) in
the rows between the potential probes. For a
square array, this procedure yields the average voltage
across one SNS junction, and this quantity is represented
by the abscissa axis in Fig.~\ref{fig3} (below, all dependences
of the differential resistance on the bias voltage
are plotted in the same coordinate systems).
The dependences of $dV/dI$--$V$ are symmetric with respect to the
direction of the current.
As one can see from the dependences
shown in Fig.~\ref{fig3}, an increase in temperature
leads to a suppression of the excess conductivity
and, at $T \simeq 570$~mK, the minimum observed at
$V = 0$ is replaced by a maximum.
It should be emphasized that
the change of a minimum to a maximum in the differential
resistance dependences at zero bias voltage occurs in
the same temperature region within which the anomaly
is observed in the temperature dependence of resistance
(in Fig.~\ref{fig2}, for sample {\bf S\#1}, this interval is indicated
by arrows).

Let us estimate the characteristic energy scales for
the two-dimensional array of SNS junctions under
study. The energy corresponding to the voltage, at
which the minimum is suppressed at zero bias and temperatures
below 570~mK, is estimated as $eV_{ex} \simeq 3$ $\mu$eV.
The presence of the maximum in the differential resistance
at $T > 570$~mK may be caused by the minimum
that occurs in the density of states in the normal region.
The theoretical study of the properties of diffusive NS
junctions~\cite{Wees} shows that the decrease in the density of
states is caused by the presence of coherent Cooper
pairs in the normal metal. According to the theory, at a
distance $x$ from the NS interface, only the pairs from the
energy window $E_F \pm E_{\text{Th}}$ (the coherence window),
where $E_{\text{Th}} = \hbar D / x^2$ is the Thouless energy, will remain
coherent. 
Therefore, the density of states has a minimum
at energies $E < E_{\text{Th}}$ and a maximum at $E \simeq E_{\text{Th}}$,
and value of the density of states in the maximum is 
higher than that in the normal metal. 
The experimental study of the density
of states as a function of the distance to the NS interface~\cite{Devoret}
showed a good agreement with the theory. Note
that this effect manifests itself in two ways. On the one
hand, the presence of coherent Cooper pairs leads to a
decrease in the resistance of the normal region (an analog
of the Maki-Thompson correction), and, on the
other hand, the pairing leads to a decrease in the density
of single-particle states (an analog of the correction to
the density of states in the Cooper channel).
These competing contributions may lead to a nonmonotonic
temperature dependence of the resistance of SNS junctions~\cite{Volkov97}.
Let us estimate the Thouless energy $E_{\text{Th}} = \hbar D / L^2$
for the structure under study, where $L = 0.2$~$\mu$m
is the length of the normal region and $D = 2$ cm$^2$/s
is the diffusion coefficient for the reference film. Then, the
Thouless energy is $E_{\text{Th}} \simeq 3$ $\mu$eV, which nearly coincides
with the characteristic voltage value (in Fig.~\ref{fig3}, this
value is indicated by the vertical straight line).
However, for the observation of the effects associated with
changes in the quasiparticle spectrum at the Thouless
energy, the necessary condition is $kT < E_{\text{Th}}$.
In the temperature range of interest, e.g., at $T = 580$~mK, the thermal
broadening is $kT \simeq 50$ $\mu$eV, which noticeably
exceeds the correlation energy $E_{\text{Th}}$ for the SNS junctions
forming the array. Thus, one should expect that the processes determined by
$E_{\text{Th}}$ would not manifest themselves in the differential resistance 
curves. However, the experimental dependences clearly display the
suppression of the maximum at a voltage of about $E_{\text{Th}}/e$.
This fact suggests that the effect observed in the
experiment is collective; i.e., the whole array is responsible
for its manifestation. Note that the total bias voltage
is $eV_{ex} \cdot 100 = eV_{\Sigma} = 300$~$\mu$eV and 
it is much higher than $kT$. 
We will return to discussing this issue after presenting
results that, in our opinion, also testify to a correlation
in the behavior of the array of SNS junctions.

\begin{figure}
\includegraphics[width=1. \columnwidth]{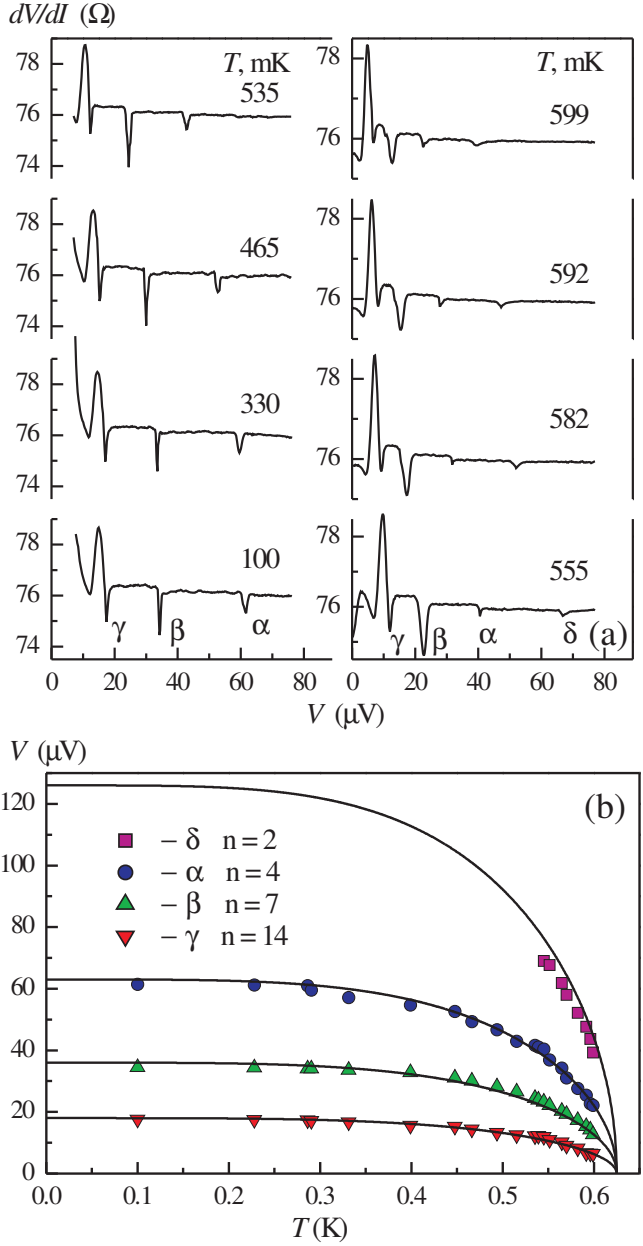}
\caption{Sample {\bf S\#1}: (a) differential resistance per square
versus the bias voltage per SNS junction (i.e., the voltage
across the potential terminals is divided by 100); (b) positions
of the features marked in Fig.~\ref{fig4}a as functions of temperature.
The solid lines represent $2 \Delta(T)/en$. The respective
values of $n$ are indicated in the plot. The dependences are
calculated for $\Delta(0) = 126$~$\mu$eV and $T_c= 0.625$~K.}
\label{fig4}
\end{figure}
Figure~\ref{fig4}a shows the dependences of the differential
resistance on the bias voltage per one SNS junction.
These dependences have pronounced minima at some
voltage values (the minima are marked by $\alpha$, $\beta$, and
$\gamma$ in the plot). When temperature increases, the minima are
shifted to lower voltages and, in the interval from 550
to 600 mK, one more minimum ($\delta$) appears; at lower
temperatures, this minimum is absent.
The temperature dependences of the positions of these features are
shown in Fig.~\ref{fig4}b by different symbols. The solid lines
in Fig.~\ref{fig4}b represent the temperature dependences of the
superconducting energy gap $2 \Delta(T)/en$, where $\Delta(T)$
is the temperature dependence of the superconducting
energy gap predicted by the BCS microscopic superconductivity
theory and $n$ is an integer (the corresponding values are indicated in 
Fig.~\ref{fig4}b). The exact value of $\Delta(0)$
for platinum monosilicide is unknown. 
The estimate by the formula $\Delta(0) = 1.76 k T_c$
($T_c = 0.88$~K for bulk PtSi~\cite{Matthias}) yields the value
$\Delta / e \simeq 133$~$\mu$V. If we set $\Delta / e = 126$~$\mu$V,
the positions of the features in the voltage
dependences of the differential resistance will be
determined by the condition $eV = 2 \Delta(T) / n$.
From Fig.~\ref{fig4}b, one can see that the temperature dependences
of the positions of minima observed in the experimental
differential resistance curves functionally coincide with
the dependences $2 \Delta(T)/en$. From this fact, we can conclude
that these features represent the subharmonic energy gap structure (SGS).

However, there are two circumstances clearly indicating
that it is difficult to treat the features observed at
bias voltages equal to $2 \Delta(T)$ as the manifestation of the
SGS caused by multiple Andreev reflections that occur
{\it independently} in each SNS junction and that this interpretation
is impossible for large values of $n$.
First, the conventional mechanism of the SGS formation in a single
SNS junction implies that, for the appearance of a subharmonic
of number $n$, an $n$-fold passage of quasiparticles
through the normal region is necessary without any
energy relaxation. In other words, the energy relaxation
length ($l_{\epsilon}$) must be no smaller than $n \cdot L$~\cite{Bez}.
The energy relaxation length can be estimated from the known
phase breaking length $l_{\varphi}$.
In the temperature range of interest, $l_{\epsilon}$ can exceed
$l_{\varphi}$ by no more than an order of magnitude~\cite{DSI}.
When a carrier acquires an energy of $2\Delta = 252$~$\mu$eV, 
it is heated to a temperature of $\sim$3 K.
According to our magnetotransport measurements, at
this temperature, we have $l_{\varphi} = 40$ nm.
Then, the most optimistic estimate yields $l_{\epsilon} \sim 0.4$~nm.
This value is much smaller than $14 \times 0.2$~$\mu$m~ $= 2.8$~$\mu$m,
which is necessary for the realization of the subharmonic with $n = 14$.
Second, it is difficult to expect that normal regions
are fully identical, i.e., characterized by exactly the
same resistance. Evidently, we deal with a network of
different resistances. Therefore, when a current flows
through the structure, the voltages across the normal
regions are different. This means that, if, for some of
the normal regions, the condition $eV = 2\Delta /n$
is satisfied at a given value of the current, for other regions it will
be satisfied at some other values of the current. Only
when the resistances are close to each other, it is possible
to observe the subharmonic energy gap structure as a result
of the statistical averaging; however, this structure will
be smeared in proportion to the deviations of actual
resistances from the average value. These speculations
lead to a natural conclusion that, for a network of random
resistances, the observation of the subharmonic energy
gap structure is impossible. Contrary to this conclusion,
the experiment shows that the SGS still manifests itself,
although it is somewhat irregular; i.e., it lacks some of
the harmonics. Similar results for an array of SNS junctions
of other configuration were obtained in~\cite{2DSNS}.
By now, no theory has been developed to describe multiply
connected SNS systems, and, on the basis of the theoretical
results obtained for single SNS junctions, it is
difficult to explain the observation of subharmonics
with numbers reaching $n = 14$.

\begin{figure}
\includegraphics[width=1. \columnwidth]{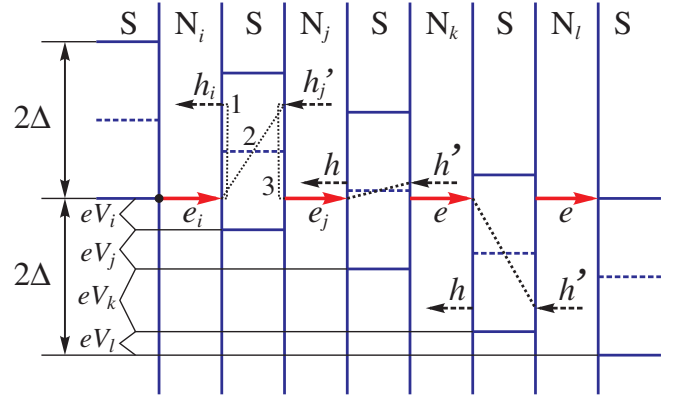}
\caption{Andreev reflection process in a semiconductor representation
for the structure consisting of alternating superconducting
and normal regions.}
\label{fig5}
\end{figure}
In considering the properties of single SNS junctions,
the superconducting regions are usually assumed
to be large and serving as reservoirs for electrons. The
fundamental distinction of the array of SNS junctions is
the finite size of the superconducting regions ($L_S$).
Therefore, it is necessary to take into account the possibility
of the mutual influence of the Andreev reflection
processes that occur at the interfaces of one superconducting
island. In NSN junctions, in addition to the
``common'' Andreev reflection, when an electron-like
excitation is reflected at the interface of the superconductor
with transformation into a hole-like excitation
(in Fig.~\ref{fig5}, process $1$: $e_i \leftrightarrow h_i$),
an additional process
of crossed Andreev reflection $e_i \leftrightarrow h_{j}^{'}$ (process $2$) is
possible, when the electron-like and hole-like excitations
are on different sides of the superconducting region~\cite{CAR1, CAR2, Yacobs}.
As a consequence, such a reflection in
combination with the common Andreev reflection,
which occurs on one side of the superconductor (process $3$), 
leads to the passage of quasiparticles without any energy loss.
Note that, for the observation of the $n$th
subharmonic of Andreev reflection, the realization of
this mechanism of charge transport does not require
that the voltage across each of the normal regions be
exactly equal to $V_n = 2\Delta /en$.
It is sufficient that the total
voltage across the chain of normal regions connected in
series through superconducting regions be equal to $2 \Delta /e$:
\begin{equation*}
\sum_{s=i}^{i+n-1} V_s = 2 \Delta /e.
\end{equation*}
Moreover, for the observation of the $n$th subharmonic
of Andreev reflection, the condition $l_{\epsilon} > nL$
needs not necessarily be satisfied. It is sufficient that the energy
relaxation length be greater than the distance between
the superconducting regions ($l_{\epsilon}>L$).
We also note that the crossed Andreev reflection process makes a considerable
contribution to the charge transport through an NSN junction even if the 
size of the superconducting region is several times greater than the superconducting
coherence length $\xi$~\cite{CAR1,CARinFSF}. In our case, $\xi(0) \simeq 70$~nm,
and at $T = 100$ mK, we have $L_S/\xi \simeq 10$.

We believe that this charge transport mechanism is
responsible for the features observed in the experiment.
Returning to the behavior of the two-dimensional array
of SNS junctions near zero bias, we note that the synchronous
Andreev reflection on both sides of the superconducting
regions should lead to correlated changes of
the energy spectra in normal regions by analogy with
the formation of minibands and minigaps in semiconductor
superlattices.

We are grateful to M.V. Feigel'man for useful discussions
and for the interest taken in our work. This
work was supported by the Russian Academy of Sciences
(the program ``Quantum Macrophysics''), the
Ministry of Science of the Russian Federation (the program
``Superconductivity of Mesoscopic and Strongly
Correlated Systems''), and the Russian Foundation for
Basic Research (project no. 03-02-16368).


\begin{thebibliography}{99}

\bibitem{Andreev} A.F. Andreev, Zh. Eksp. Teor. Fiz. \textbf{46}, 1823 
(1964) [Sov. Phys. JETP \textbf{19}, 1228 (1964)].

\bibitem{KBT} T.M.~Klapwijk, G.E.~Blonder, and M.~Tinknam,
Physica \textbf{109-110B+C}, 1657 (1982).

\bibitem{Pannetier} B. Pannetier and H. Courtois, J. of Low
Temp. Phys. \textbf{118}, 599 (2000).

\bibitem{BTK}
G.E.~Blonder, M.~Tinkham, and T.M.~Klapwijk, Phys.
Rev. \textbf{B25}, 4515 (1983).

\bibitem{OTBK}
M.~Octavio, M.~Tinkham, G.E.~Blonder, and T.M.~Klapwijk, 
Phys. Rev. \textbf{B27}, 6739 (1983); 
K.~Flensberg, J.~Bindslev Hansen, M.~Octavio, {\it ibid.} {\bf 38},
8707 (1988).

\bibitem{Bez} E.\,V.~Bezuglyi, E.\,N.~Bratus', V.\,S.~Shumeiko, 
G.~Wendin, and H.~Takayanagi, 
Phys. Rev. \textbf{B62}, 14439 (2000).

\bibitem{Q_SMT} M.\,V.~Feigel'man, A.\,I.~Larkin, and M.\,A.~Skvortsov,
Phys. Rev. Lett. \textbf{86}, 1869 (2001).

\bibitem{PtSitech} R.\,A.~Donaton, S.~Jin, H.~Bender, et al,
Microelectronic Engineering {\bf 37/38}, 504 (1997).

\bibitem{Wees} B.\,J.~van~Wees, P.~de~Vries, P.\,H.\,C.~Magnee,
and T.~M.~Klapwijk,
Phys. Rev. Lett. \textbf{69}, 510 (1992). 

\bibitem{Devoret} S.~Gueron, Norman O.~Birge, D.~Esteve, 
and M.\,H.~Devoret, 
Phys. Rev. Lett. \textbf{77}, 3025 (1996).

\bibitem{Volkov97} Anatoly~F.~Volkov and Hideaki Takayanagi.
Phys. Rev. B \textbf{56}, 11184 (1997).

\bibitem{Matthias} B.\,T.\,Matthias, T.\,H.\,Geballe, and V.\,B.\,Compton.
Rev. Mod. Phys. \textbf{35}, 1 (1963). 

\bibitem{DSI} S.\,I.~Dorozhkin, F.~Lell and W.\,Schoepe,
Solid State Comm. \textbf{60}, 245 (1986).

\bibitem{2DSNS} T.\,I.~Baturina, Z.\,D.~Kvon, and A.\,E.~Plotnikov,
Phys. Rev. B \textbf{63}, 180503(R) (2001).

\bibitem{CAR1} J.\,M.~Byers and M.\,E.~Flatt\'{e}, 
Phys. Rev. Lett. \textbf{74}, 306 (1995).

\bibitem{CAR2} G.\,Deutscher and D.\,Feinberg, 
Appl. Phys. Lett. \textbf{76}, 487 (2000).

\bibitem{Yacobs} Arne~Jacobs and Reiner K\"{u}mmel,
Phys. Rev. B \textbf{64}, 104515 (2001).

\bibitem{CARinFSF} D.~Beckman, H.\,B.~Weber, H.\,v.\,L\"{o}hneysen,
Phys. Rev. Lett. \textbf{93}, 197003 (2004).

\end{thebibliography}
\end{document}